\documentclass[%
 reprint,
amsmath,amssymb,
 aps,
]{revtex4-2}

\usepackage{graphicx}
\usepackage{dcolumn}
\usepackage{bm}
\usepackage{physics}
\usepackage{mathrsfs}
\usepackage{hyperref}
\usepackage{xcolor}
\begin{document}

\preprint{APS/123-QED}

\title{Many-body dynamics with explicitly time-dependent neural quantum states}

\author{Anka Van de Walle$^{1,2}$, Markus Schmitt$^{3,4}$, and Annabelle Bohrdt$^{2,4}$\footnote{Corresponding author email: annabelle.bohrdt@ur.de}} 

\address{$^{1}$Ludwig-Maximilians-University Munich, Theresienstr. 37, Munich D-80333, Germany}
\address{$^{2}$Munich Center for Quantum Science and Technology, Schellingstr. 4, Munich D-80799, Germany}
\address{$^{3}$Institute of Quantum Control (PGI-8), Forschungszentrum Jülich, D-52425 Jülich, Germany}
\address{$^{4}$University of Regensburg, Universitätsstr. 31, Regensburg D-93053, Germany}

\date{\today}

\begin{abstract}
Simulating the dynamics of many-body quantum systems is a significant challenge, especially in higher dimensions where entanglement grows rapidly. Neural quantum states (NQS) offer a promising tool for representing quantum wavefunctions, but their application to time evolution faces scaling challenges. We introduce the time-dependent neural quantum state (t-NQS), a novel approach incorporating explicit time dependence into the neural network ansatz. This framework optimizes a single, time-independent set of parameters to solve the time-dependent Schr\"odinger equation across an entire time interval. We detail an autoregressive, attention-based transformer architecture and techniques for extending the model's applicability. To benchmark and demonstrate our method, we simulate quench dynamics in the 2D transverse field Ising model and the time-dependent preparation of the 2D antiferromagnetic state in a Heisenberg model, demonstrating state of the art performance, scalability, and extrapolation to unseen intervals. These results establish t-NQS as a powerful framework for exploring quantum dynamics in strongly correlated systems.
\end{abstract}

\maketitle

\begin{figure*}[t]
\includegraphics[clip, trim=3cm 1.4cm 0.5cm 3cm, width=18cm]{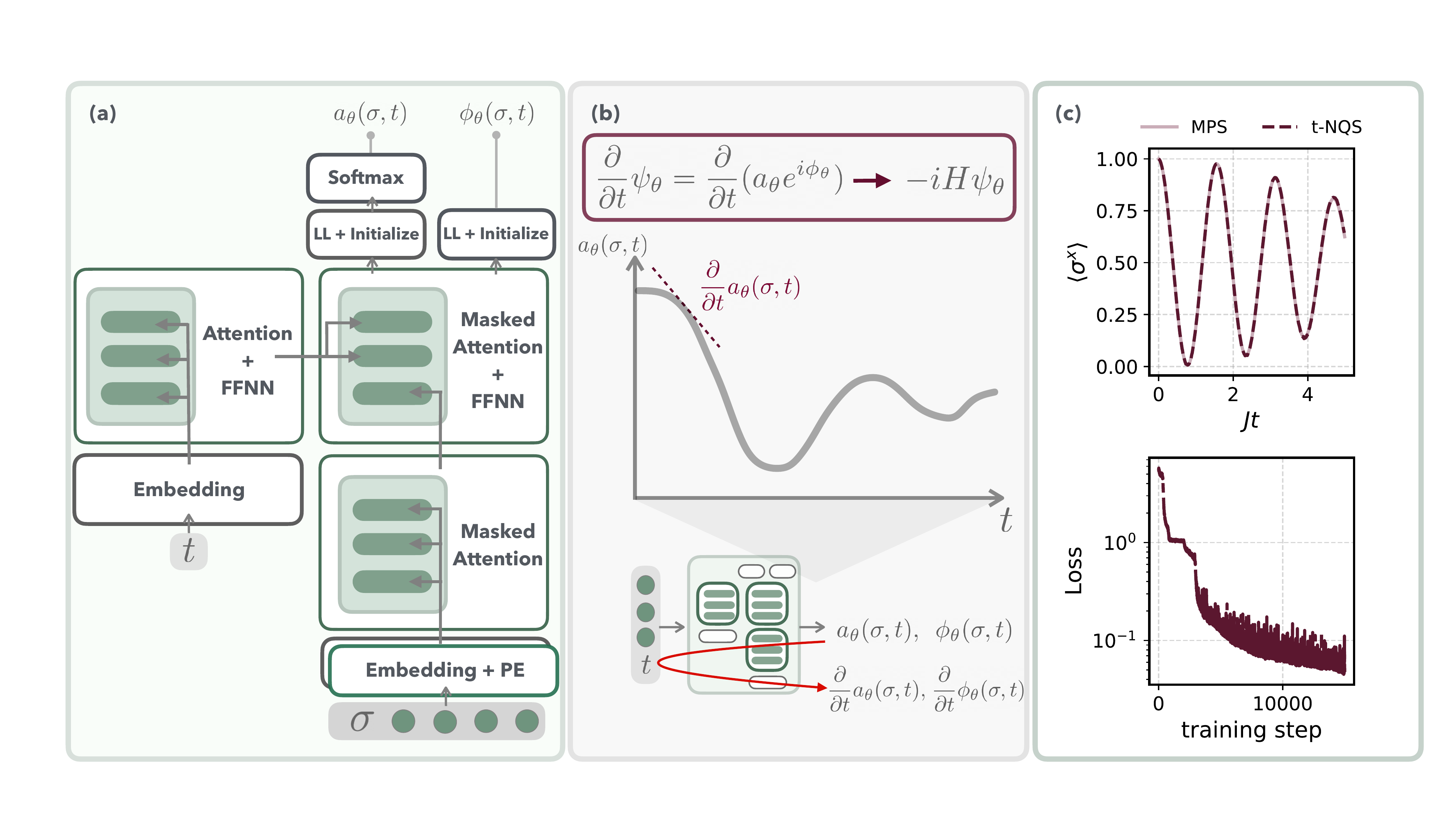}
\centering
\caption{  
(a) Schematic representation of the transformer model of the t-NQS with an encoder-decoder structure. The encoder captures the time dependency, while the decoder processes the spin configuration, producing the amplitude and phase of the quantum state as outputs.  
(b) Sketch of the dynamics captured by the t-NQS model, where the model is optimized by minimizing the error in the time-dependent Schrödinger equation. The time derivative of the wavefunction is computed efficiently using backpropagation.
(c) Results for a quench of the 1D TFI model with $N=40$ sites and periodic boundary conditions, including the dynamics captured by the trained network and the loss (i.e. the evaluation of the cost function) during model optimization. The quench is performed with $h/h_c = 0.1$ on the initial state $\ket{\psi_0} = \ket{\rightarrow,...,\rightarrow}$.}
\label{fig:mainFig}
\end{figure*}


Experimental and theoretical advancements have in recent years revealed a plethora of intriguing non-equilibrium phenomena, such as pre-thermalization \cite{langen2016prethermalization}, many-body localization \cite{abanin2017recent}, dynamical quantum phase transitions \cite{heyl2018dynamical, diehl2010dynamical}, time crystals \cite{zhang2017observation,Choi2017}, or quantum scars \cite{bernien2017probing}.
The expanding capabilities of quantum simulators are now turning a spotlight on two-dimensional systems in particular \cite{Hild2014,Bordia2017,Scholl2021,Ebadi2021,Manovitz2024}, highlighting the fact that the development of effective theoretical tools to address this regime is still a cornerstone challenge in quantum many-body physics.
%
%
Despite their importance, capturing real-time dynamics remains a difficult task, particularly due to the exponential growth of the Hilbert space and the intricate entanglement generated over time. This renders exact simulations computationally intractable even for modestly sized systems. Tensor network (TN) methods have made remarkable strides in simulating one-dimensional systems \cite{Paeckel2019}, leveraging their ability to efficiently represent low-entanglement states. However, their extension to two-dimensional dynamics faces significant challenges due to the rapid growth of entanglement, even in systems governed by an area law, or the intractable complexity of tensor contractions.   

Neural quantum states (NQS) provide a promising alternative, offering the flexibility to encode complex correlations through variational wavefunctions parameterized by artificial neural networks (ANNs) \cite{deng2017quantum}. These models have been successfully applied to ground-state calculations \cite{bukov2021learning, nomura2021helping, viteritti2023transformer,  lange2024architectures}, real-time dynamics \cite{schmitt2020quantum,Schmitt2022,Medvidovic2023,MendesSantos2023,MendesSantos2024, nys2024ab}, finite-temperature simulations \cite{nys2024real}, and open-system dynamics \cite{Hartmann2019,Reh2021,luo2022autoregressive}, providing a complementary approach to tensor networks in higher dimensions or high-entanglement regimes.
While the expressivity of NQS is enhanced as compared to TNs \cite{Deng2017,Levine2019,Sharir2022}, their effective optimization appears as one major bottleneck in applications \cite{Bukov2021,Chen2024,Kolnamer2024}, especially in the time-dependent setting \cite{schmitt2020quantum,Hofmann2022,donatella2023dynamics,sinibaldi2023unbiasing}.  

For isolated systems, the non-equilibrium dynamics of interest is described by the time-dependent Schrödinger equation. 
Since it is a first-order ordinary differential equation, numerical solutions are usually obtained by forward integration of an initial value problem with discrete time steps.
A common strategy for both TNs and NQS is to employ a time-dependent variational principle (TDVP) in order to project Schrödinger's equation onto the reduced state space defined by the ansatz \cite{PhysRevB.94.165116, carleo2017solving, schmitt2020quantum}. 
In the case of NQS, the TDVP is solved by a variety of time-dependent variational Monte Carlo (tVMC) methods \cite{carleo2017solving, schmitt2020quantum, Gutierrez2022, Medvidovic2023, Nysvariational,sinibaldi2023unbiasing,donatella2023dynamics}.
Although this approach has enabled the numerous impressive applications mentioned above, the numerical forward-propagation poses significant challenges.
Solving the TDVP in parameter space means integrating a high-dimensional non-linear dynamics equation, that may even become stiff. 
Highly accurate solutions are required at every time step in order to keep the accumulation of errors under control until late times.
This is particularly challenging because existing algorithms rely on probing the current wave function at each time using Monte Carlo sampling. Thus, the resulting statistical noise restricts the accurately simulable time scales. 
Moreover, the forward integration scheme is inherently sequential, limiting the possibilities to leverage parallel computing resources.  

In this context, we introduce the explicitly time-dependent neural quantum state (t-NQS) as a fundamentally different method to simulate quantum many-body dynamics. 
Unlike conventional time-evolution techniques, t-NQS eschews forward-integration schemes. 
Instead, we recast the problem of solving the time-dependent Schrödinger equation as a global optimization task on a targeted time interval, crucially enabled by automatic differentiation techniques \cite{Baydin2018}. 
In combination with the expressive power of ANNs, this paradigm shift allows for the simultaneous optimization of the wavefunction on a whole time interval. 
%
%
This enables scalable and efficient simulations of dynamics in regimes that were previously computationally prohibitive. 
We demonstrate accurate simulations of the quench dynamics of a two-dimensional quantum magnet, which can be naturally emulated using Rydberg atom quantum simulators. The precision of these results can be assessed in a self-sufficient manner and errors can be systematically reduced by increasing the ANN size. We furthermore report the remarkable ability of the t-NQS to extrapolate to unseen time intervals.
%


\section{Global variational principle}

\subsection{Variational objective}
We will be interested in solving the time-dependent Schrödinger equation
\begin{align}
    i\frac{d}{dt}\ket{\psi}=H\ket{\psi}\ ,
    \label{eq:seq}
\end{align}
which describes the dynamics of isolated quantum systems as it is, for example, realized in present-day quantum simulators. For a many-body system the Hilbert space is spanned by computational basis states $\{ \ket{\sigma}\}$, where $\sigma\equiv(\sigma_1,\ldots,\sigma_N)$ represents the local quantum numbers for a system of size $N$, such as Fock occupation numbers or local spin configurations. While our approach is more general, we will from now on consider qubit degrees of freedom, i.e. $\sigma_i\in\{0,1\}$. Upon expanding in the chosen basis, $\ket{\psi(t)}=\sum_\sigma \psi_\sigma(t)\ket{\sigma}$, the state $\ket{\psi}$ is fully described by the set of time-dependent wave function coefficients $\psi_\sigma(t)$.
Clearly, solutions of Eq.~\eqref{eq:seq} for the full wave function are restricted to limited system sizes due to the exponential growth of the Hilbert space dimension with the system size.

Instead of explicitly keeping track of all wave function coefficients, the idea of NQS is to find compressed representations of the wave function in the form of artificial neural networks \cite{carleo2017solving}.
Concretely, an NQS constitutes a parameterized function $\psi_\theta(\sigma)$ -- the artificial neural network -- with variational parameters $\theta$.
This function constitutes a compressed wave function representation, if one can find an efficient parameterization, such that $\psi_\theta(\sigma)\approx\psi_\sigma$ for a given target wave function $\psi_\sigma$.
In all state-of-the-art approaches for time evolution with NQS, the time-dependence enters via time-dependent parameters $\theta(t)$, i.e., $\ket{\psi(t)}=\sum_\sigma\psi_{\theta(t)}(\sigma)\ket{\sigma}$, and the parameters are propagated iteratively by solving a TDVP \cite{carleo2017solving, schmitt2020quantum, Gutierrez2022, Medvidovic2023, Nysvariational,sinibaldi2023unbiasing,donatella2023dynamics}.
By contrast, we consider an explicitly time-dependent NQS $\psi_\theta(\sigma, t)$ with the goal to optimize the \emph{time-independent} parameters $\theta$, such that $\psi_\theta(\sigma, t)$ solves the Schrödinger equation \eqref{eq:seq} on a whole time interval $t\in[0,T]$.
For this purpose, we formulate a variational objective, which is minimized by solutions of the Schrödinger equation.
In McLachlan's formulation \cite{McLachlan1964,broeckhove1988equivalence}, the TDVP amounts to minimizing the residual $\mathcal L_\mathrm{McL}(\dot\theta)=||\frac{d}{dt}\ket{\psi_\theta}+iH\ket{\psi}||=||\sum_k\dot\theta_k\frac{\partial}{\partial\theta_k}\ket{\psi_\theta}+iH\ket{\psi}||$ to find the locally optimal rate of change $\dot\theta\equiv\frac{d}{dt}\theta$ of the time-dependent parameters.
In order to minimize the deviation from Schrödingers equation across a whole time interval with time-independent parameters, we consider the integral of the time-local residuals, and formulate a cost function as
\begin{align}
    \mathcal L(\theta)&=\frac{1}{T}\int_{0}^{T}  \varepsilon_\theta(t) \,dt 
    \nonumber\\
    &= \frac{1}{T}\int_{0}^{T}  \big|\frac{d}{dt} \ket{\psi_\theta(t)} + i H \ket{\psi_\theta(t)} \big|^2 \,dt\ .
    \label{eq:integral_cost}
\end{align}
Notice, that in analogy to the conventional TDVP, this cost function could be altered to be insensitive to the normalization and global phase of $\ket{\psi_\theta(t)}$.

In practice, we approximate the integral in Eq.~\eqref{eq:integral_cost} as a discrete sum over time steps $\{t_i\}_i$, which are evenly distributed within the training interval. The spacing between consecutive target time steps is denoted as $\Delta t = t_{i+1} - t_i$, chosen to be sufficiently small to ensure an accurate approximation of the integral as
\begin{align}
    \mathcal L(\theta)=\frac{\Delta t}{T}\sum_{t_i} \varepsilon_\theta(t_i)\ .
    \label{eq:cost_function}
\end{align}
In the usual manner of variational Monte Carlo (VMC) \cite{becca2017quantum}, the time-local residuals can be rewritten in the form of an expectation value with respect to the Born distribution $p_\theta(\sigma,t) = |\psi_\theta(\sigma,t)|^2$,
\begin{align}
    \varepsilon_\theta(t)=\sum_\sigma p_\theta(\sigma,t)\varepsilon_\mathrm{loc}^\theta(\sigma,t)\ ,
\end{align}
where we introduced the local estimator of the residual,
\begin{align}
    \varepsilon_\mathrm{loc}^\theta(\sigma,t)=\Big|\frac{d}{dt}\log\big(\psi_\theta(\sigma, t)\big) + i E_\mathrm{loc}^\theta(\sigma, t)\Big|^2
\end{align}
and the local energy $E_\mathrm{loc}^\theta(\sigma, t_i) = \frac{\bra{\sigma}\hat{H} \ket{\psi_\theta(t_i)}}{\braket{\sigma}{\psi_\theta(t_i)}}$. Therefore, the cost function \eqref{eq:cost_function} can be efficiently estimated by sampling the Born distribution using standard techniques of VMC.
We note that similar to tVMC \cite{sinibaldi2023unbiasing}, this formulation relies on positivity of the Born distribution, $p_\theta(\sigma,t)>0$, but this limitation did not appear relevant within our study.  

A crucial point in this approach is the ability to evaluate the time derivative $\frac{d}{dt}\log\big(\psi_\theta(\sigma, t)\big)$. For this purpose, we construct the NQS ansatz such that the time $t$ is an additional input variable besides the computational basis configuration $\sigma$. Computing the time derivative is then facilitated in an efficient and exact manner by automatic differentiation \cite{Baydin2018}. 

Using this global variational objective, we employ gradient-based optimization in order to find the NQS parameters $\theta$, that minimize the accumulated residual. Gradients of the cost function are obtained by automatic differentiation and the problem is amenable to leveraging the established optimization toolbox of deep learning \cite{Sun2020}, see Methods for further details. The t-NQS approach is schematically illustrated in Fig.\ref{fig:mainFig}.

We furthermore introduce the time-dependent accumulated residuals as a figure of merit indicating the quality of the obtained results:
\begin{align}
    \mathcal R_\theta(t)=\frac{\Delta t}{T}\sum_{t_i\le t} \varepsilon_\theta(t_i)\ .
    \label{eq:accumulated_residuals}
\end{align}
We will discuss in the following Section, how the initial condition $\ket{\psi(t=0)}=\ket{\psi_0}$ can be exactly fixed within our approach. This means, that the inaccuracy will increase with $t$ and we will demonstrate in Section \ref{subsec:accuracy} that the accumulated residuals \eqref{eq:accumulated_residuals} are suited to assert the quality of the obtained result.

\begin{figure}[b]
\includegraphics[clip, trim=11.5cm 4cm 15.5cm 6cm, width=8.5cm]{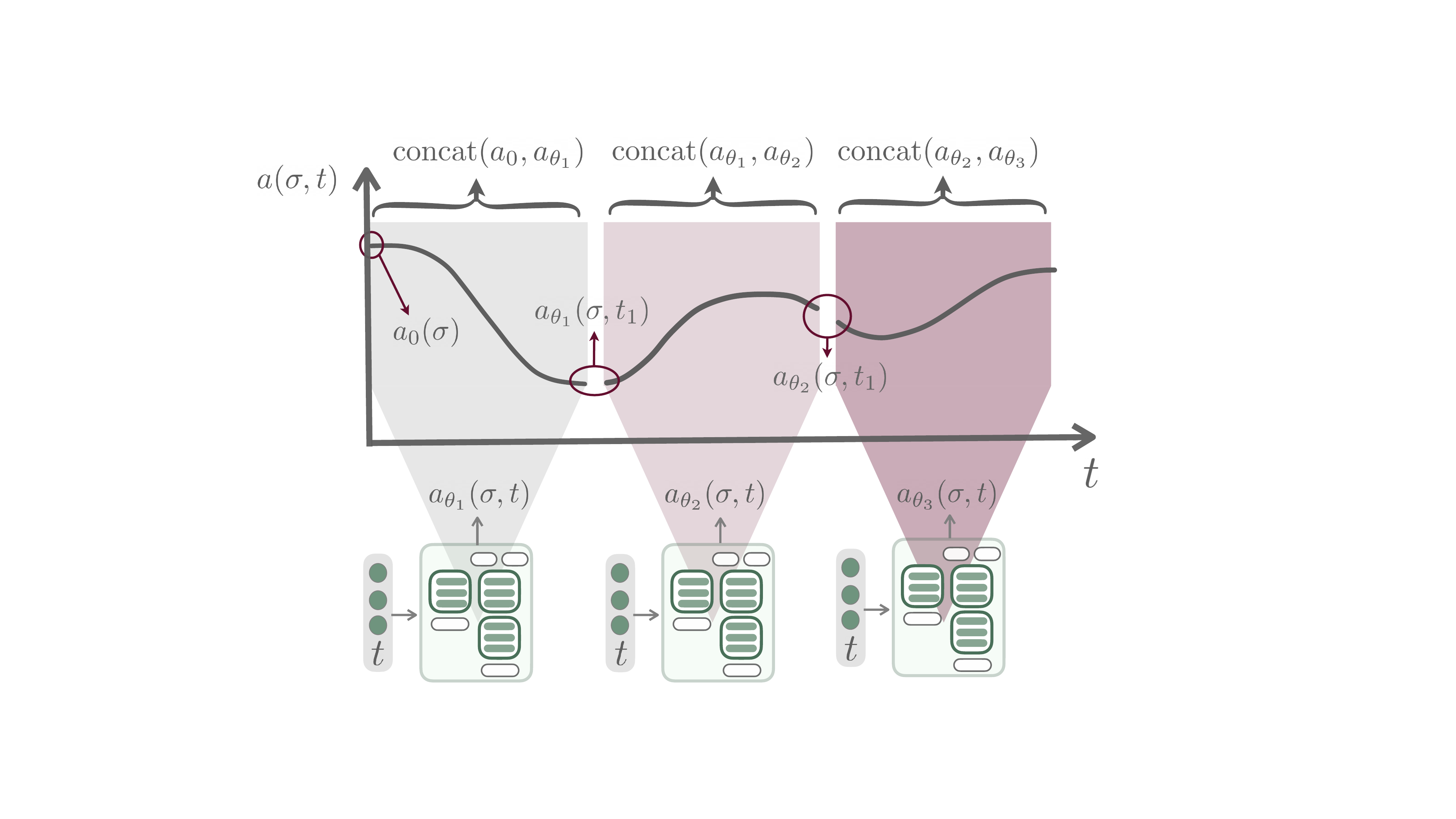}
\centering
\caption{Schematic representation of the dynamics captured by the t-NQS model. The figure illustrates how the model represents the quantum state across all time steps within a single time interval and demonstrates the concatenation of multiple t-NQS models to extend the simulation across larger time scales.}
\label{fig:concat}
\end{figure}

\begin{figure*}[t]
\includegraphics[clip, trim=0.2cm 12cm 0.15cm 3cm, width=18cm]{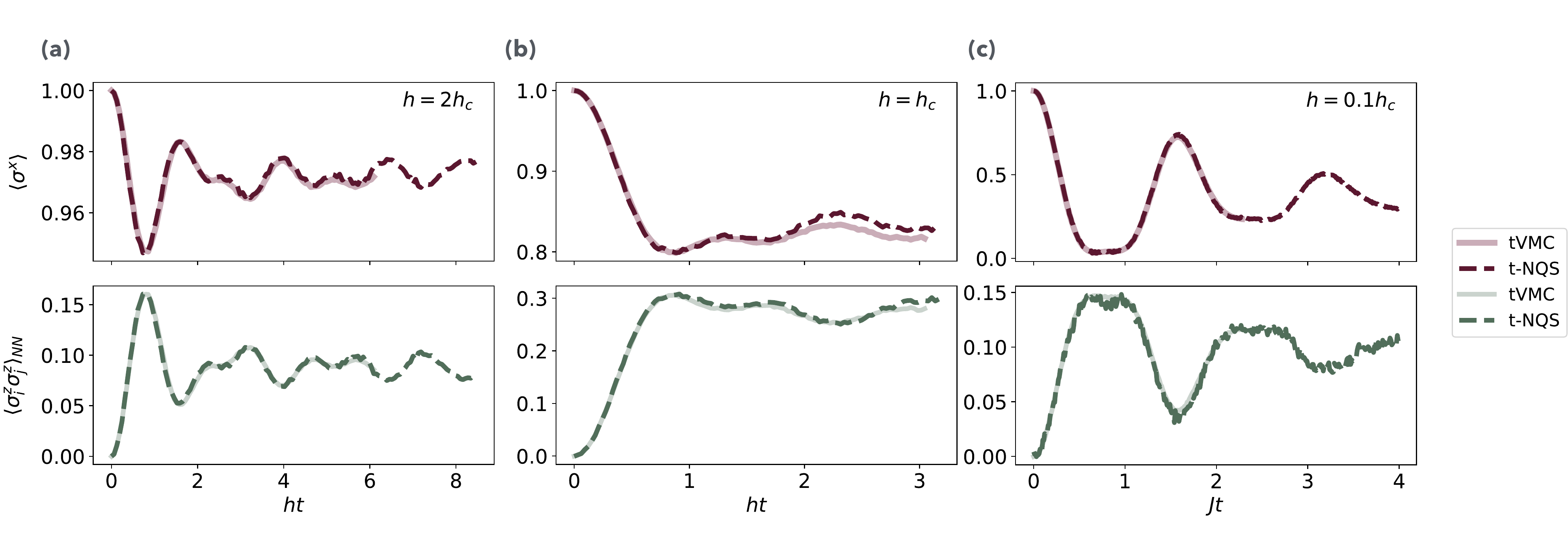}
\centering
\caption{Time evolution of the transverse spin polarization $\langle \sigma^x \rangle = \frac{1}{N} \sum_i \langle \sigma_i^x \rangle$ after quenching the 2D TFI model ($N = 8 \times 8$ with PBC) from the paramagnetic initial state $\ket{\psi_0} = \ket{\rightarrow,...,\rightarrow}$: (a) into the paramagnetic phase at $h = 2h_c$, (b) to the critical point, and (c) into the ferromagnetic phase at $h = 0.1 h_c$.  The second row displays corresponding correlation functions $\langle \sigma_{i}^z \sigma_{j}^z \rangle_{\text{NN}} = \frac{1}{N} \sum_{\langle i, j \rangle} \langle \sigma_{i}^z \sigma_{j}^z \rangle $ between all nearest neighboring sites $\langle i, j \rangle$. The results are compared to data obtained with tVMC using NQS from Ref. \cite{schmitt2020quantum}, showing good agreement across all cases. }
\label{fig:results}
\end{figure*}

\subsection{Initial condition}
Given that the t-NQS model represents the quantum state at every time step, it is essential to establish a method to initialize the model at $t = t_0$. We do this by incorporating an initialization layer within the neural network architecture, at the end of the network, which serves to effectively blend the fixed initial state $\psi_0(\sigma) = a_0(\sigma) e^{i\phi_0(\sigma)}$ with the outputs of the neural network. This layer performs a linear combination of $\psi_0(\sigma)$ and the network's output $\psi_\theta(\sigma, t)$, after the final trainable layer. By doing so, we ensure that the model starts from a well-defined initial state, while still allowing the neural network to adapt this state for all later time steps based on the learned dynamics of the system. This initial state can be any well-defined quantum state for which $ a_0(\sigma)e^{i\phi_0(\sigma)}$ is known -- for instanceas an NQS. 
Concretely, we opt for a simple time-dependent linear combination, which blends the initial amplitude of the state with the amplitude at all later time steps, concatenating the different NQS models:
\begin{equation} \label{eq:concat}
    \text{concat}(a_0, a_{\theta_1}) = f(t) a_0(\sigma) + (1-f(t)) a_{\theta_1}(\sigma, t) 
\end{equation}
This blending is also applied to the phase of the state in an entirely analogous manner. By selecting a differentiable function $f(t)$ that meets the criteria $f(t_0) = 1$ and $f(t_1) = 0$, we can effectively initialize the network at the first time step while simultaneously ensuring that the initial state has no influence at the final time step.   

The concatenation method enables the use of multiple t-NQS models to describe the dynamics, alleviating the need for a single model with an increasingly large number of parameters to describe longer time scales. Consecutive t-NQS models can capture the system’s dynamics by dividing the evolution into smaller intervals, each labeled $[t_0, t_1]$, with each new model initialized at the final time step of the previous one, see Fig.\ref{fig:concat}.

\section{Transformer t-NQS Architecture}
For the underlying architecture of the t-NQS model, we employ the encoder-decoder structure of a transformer model, as proposed in Ref.~\cite{vaswani2017attention}. 
Given their prior use in incorporating context such as Hamiltonian parameters and system size \cite{fitzek2024rydberggpt, zhang2023transformer}, transformer models naturally extend to capturing time dependency. These generative models are universal sequence-to-sequence function approximators that, with sufficiently many tunable parameters $\theta$, can encode arbitrarily complex functions and naturally capture all-to-all interactions. The structure of the transformer architecture used for the t-NQS model is depicted in Fig. \ref{fig:mainFig}a. We employ an encoder-decoder structure of the transformer model, where the encoder embeds the context of the quantum state, here the time input parameter, into an embedding space of a higher dimension $d_e$. We feed the embedded vector to a plain transformer network, consisting of an attention layer and a feed forward neural network, where the model captures correlations within the input and forms a context vector. This is in turn passed to the decoder structure of the transformer. The decoder fulfills a double purpose, both encoding the discrete variables representing the physical degrees of freedom, such as the spin or the occupation number of the system, while combining this with the encoded context vector. Secondly, it decodes this information into both an amplitude and a phase function, effectively modeling the quantum state $\ket{\psi(t)}$.  

For the decoder we adopt an autoregressive model, which takes as an input $N$ tokens, in this case the spin state $\sigma = \{\sigma_i\}_{i=1}^N$, and outputs both the normalized probability distribution $p_\theta(\sigma, t) = |\psi_\theta(\sigma, t)|^2 = a_\theta(\sigma, t)^2$, corresponding to the state's amplitude, and the phase $\phi_\theta(\sigma, t)$. Note that the t-NQS framework is not restricted to transformer or autoregressive architectures; it can also be implemented using other neural network models to represent the time-dependent quantum state.  

\section{Quench and ramp dynamics}

\subsection{Transverse Field Ising quench}

We demonstrate the effectiveness of our method by simulating the dynamics of a transverse-field Ising model (TFIM) on a 2D square lattice, described by the Hamiltonian
\begin{equation}
    H = -J\sum_{\langle i,j \rangle} \sigma_i^z \sigma_{j}^z - h \sum_i \sigma_i^x\ ,
\end{equation}
where $\sigma_i^x$ and $\sigma_i^z$ are the Pauli $x$ and $z$ operators acting on site $i$, while $\langle i,j\rangle$ is the set of all nearest neighboring sites in the lattice. Here, $J$ is the coupling strength and $h$ is the external field strength. This model is characterized by a quantum phase transition at the critical external field strength $h_c/J = 1$ in 1D, and $h_c/J = 3.04438$ in 2D.
It is of interest as it has successfully been implemented on quantum simulators using 2D ion traps \cite{guo2024site}, as well as on Rydberg atom quantum simulators \cite{labuhn2016tunable, guardado2018probing, schauss2018quantum, keesling2019quantum, browaeys2020many}, and acts as an important benchmark for dynamics simulation using NQS \cite{carleo2017solving, schmitt2020quantum, donatella2023dynamics, sinibaldi2023unbiasing, Gravina2024}.  

We prepare the system in the ground state of the TFIM Hamiltonian for $h \gg J$, namely the paramagnetic ground state given by $\ket{\psi_0} = \ket{\rightarrow,...,\rightarrow}$, and quench the magnetic field to the values $h/h_c = 2, 1,$ and $0.1$. In our simulations, we consider periodic boundary conditions (PBC). A quench near the critical point is anticipated to be the most challenging to simulate, as many excitations are generated due to the vanishing energy gap, and entanglement grows particularly rapidly. 
This makes it difficult to find efficient representations of the quantum state.  

The dynamics of the transverse magnetization $\langle\sigma^x\rangle=\frac1N\sum_i\langle\sigma_i^x\rangle$ and the longitudinal correlation $\langle\sigma_i^z\sigma_j^z\rangle_\mathrm{NN}=\frac1N\sum_{\langle i,j\rangle}\langle\sigma_i^z\sigma_j^z\rangle$ of nearest neighbors $i$ and $j$ for a system of $8 \times 8$ sites are shown in Fig.~\ref{fig:results}.  We compare our results with data obtained from NQS simulations using the tVMC algorithm \cite{schmitt2020quantum}. 
For the quench within the paramagnetic phase shown in Fig.~\ref{fig:results}a, we find very good agreement with the reference data up to the latest times.
However, the build-up of correlations remains rather limited due to the weakness of the quench.
Stronger correlations are built up in the quench to the critical point, Fig.~\ref{fig:results}b. The correlation function obtained with t-NQS accurately reproduces the tVMC result, but slight deviations become apparent in the transverse magnetization at the later times.
We attribute this deviation to the limited expressivity of the t-NQS used here. Data demonstrating the possibility to systematically improve the accuracy of simulations by increasing the network size (or, equivalently, decreasing the time interval length) will be presented below in Section \ref{subsec:accuracy}.
Finally, for the quench into the ferromagnetic phase shown in Fig.~\ref{fig:results}c, we again find excellent agreement between the two methods up to the latest times reached with the tVMC reference method.
The time reached with the t-NQS method is in fact almost twice the time up to which the tVMC NQS approach could be converged and alternative numerical approaches have been limited to similar time scales \cite{Czarnik2019,Richter2020}.


\begin{figure}[t]
\includegraphics[clip, trim=12cm 5cm 16cm 5cm, width=8.5cm]{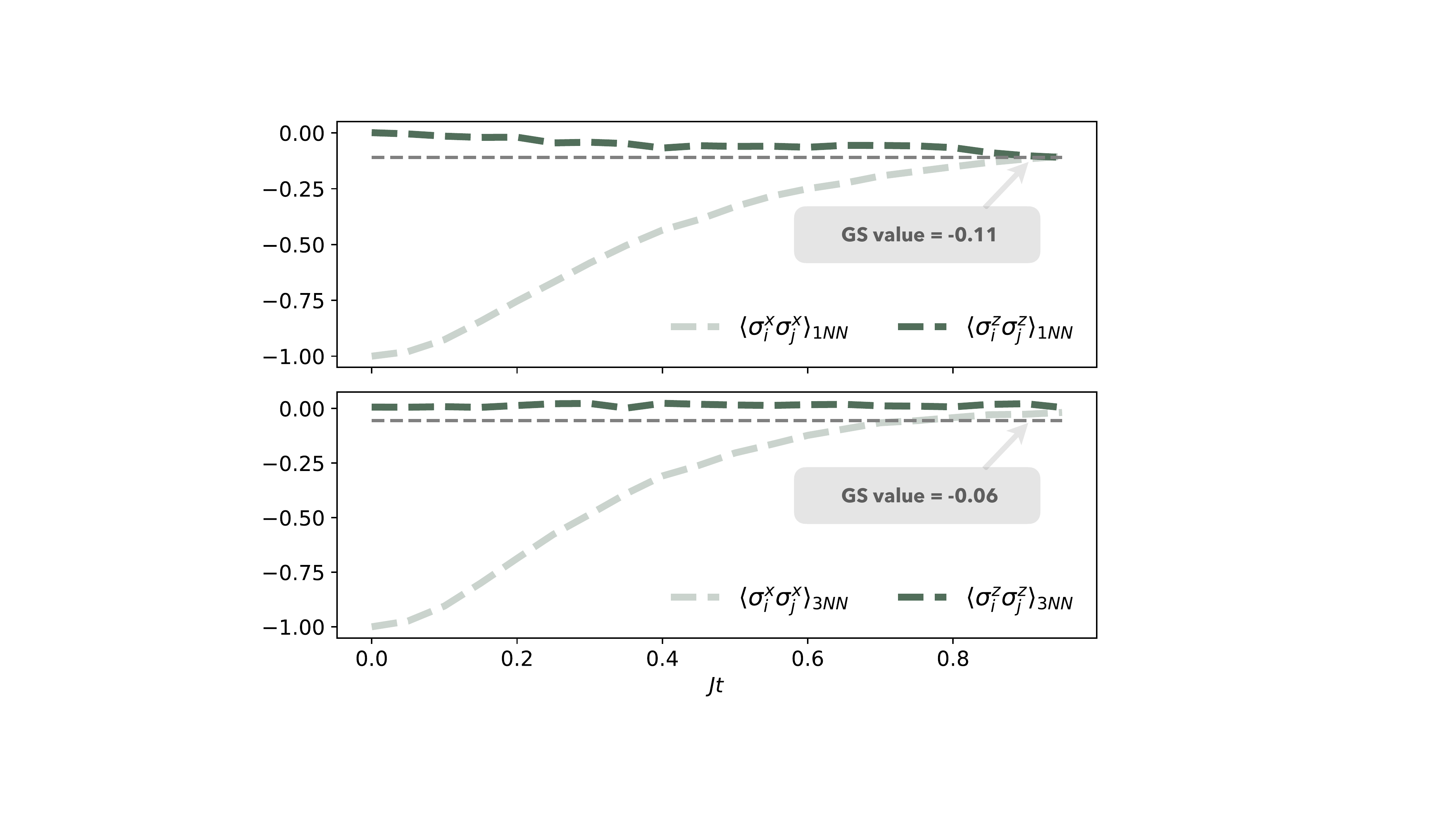}
\centering
\caption{Quench dynamics of the 2D Heisenberg model with a tuned staggered external field (from $ q(t_0) = 3 $ to  $q(t_1) = 0$ ) on a $8\times8$ lattice, initialized in the Néel state $\ket{\psi_0} = \ket{\rightarrow, \leftarrow, \rightarrow, ...}$. The upper plot shows nearest-neighbor correlations, while the lower plot presents 3-nearest neighbor correlations.  }
\label{fig:Heisenberg}
\end{figure}

\subsection{Heisenberg Antiferromagnet preparation}

We further demonstrate the capability of the t-NQS method by applying it to simulate the ramping dynamics with a time-dependent Hamiltonian. Specifically, the t-NQS approach removes the reliance on traditional time integration methods, streamlining the description of dynamics governed by time-dependent Hamiltonians. We consider the 2D Heisenberg model on a square lattice with a staggered external field, described by the Hamiltonian, 
\begin{equation}
    H(t) = J \sum_{\langle i,j \rangle} \boldsymbol{\sigma}_i \cdot \boldsymbol{\sigma}_j + q(t) \sum_{i_x,i_y} (-1)^{i_x + i_y} \sigma_i^x,
\end{equation}
where $\boldsymbol{\sigma}_i = \{ \sigma_i^x, \sigma_i^y, \sigma_i^z \}$ are the Pauli operators, $\langle i,j \rangle$ denotes all nearest-neighbor pairs on the lattice, and $i_{x,y}$ are the coordinates of site $i$.  The system transitions from a Néel state at high field to a 2D antiferromagnet (AFM) at zero field. The role of staggered fields in preparing antiferromagnetic order in 2D quantum systems has been demonstrated both numerically \cite{xie2022bayesian} and experimentally with cold-atom quantum simulators \cite{bohrdt2024microscopy}. For demonstration, we initialize the model in the ground state of the staggered external field, the Néel state $\ket{\rightarrow, \leftarrow, \rightarrow, ...}$, and quench the external field by linearly reducing $q(t)$ from $q(t_0) = 3$ to $q(t_1) = 0$, with the coupling constant of the Heisenberg term fixed at $J = 0.5$. To characterize the dynamics, we analyze the spin-spin correlations, defined as $\langle \sigma_i \sigma_j \rangle_{k\text{NN}} = \frac{1}{N} \sum_{\langle i,j \rangle_k} \langle \sigma_i \sigma_j \rangle$, where $\langle i,j \rangle_k$ represents all $k$-nearest-neighbor pairs. Here, we present the results for both short- and long-range correlations, focusing on sets of nearest neighbors (1NN) and sets of 3-nearest neighbors (3NN), where the distance between 3NN sites is 3 lattice spacings.  

The results of the time-dependent AFM preparation demonstrate that the t-NQS effectively captures the dynamics of the system, even with a time-dependent Hamiltonian. The nearest-neighbor (1NN) correlations start correctly from the initial Néel state and evolve smoothly, converging to the ground state values of the Heisenberg model as the staggered field vanishes, in agreement with the ground state results reported in Ref.~\cite{richter2010spin}. The 3-nearest-neighbor (3NN) correlations don’t agree with ground state Heisenberg values, potentially due to the ramp duration. Extended Numerical investigation of these ramps can thus provide valuable assistance to cold atom experiments aiming to prepare low temperature equilibrium states.

\subsection{Systematic enhancement of accuracy}
\label{subsec:accuracy}
Since artificial neural networks are universal function approximators in the limit of large network size \cite{Cybenko1989, Hornik1991}, it is expected, that the accuracy of solutions found with the t-NQS approach can be systematically improved by increasing the ansatz. We will now present data supporting, that this expectation indeed holds in practice.  

\begin{figure}[!]
\includegraphics[clip, trim=5.5cm 1cm 0.5cm 0.4cm, width=8.5cm]{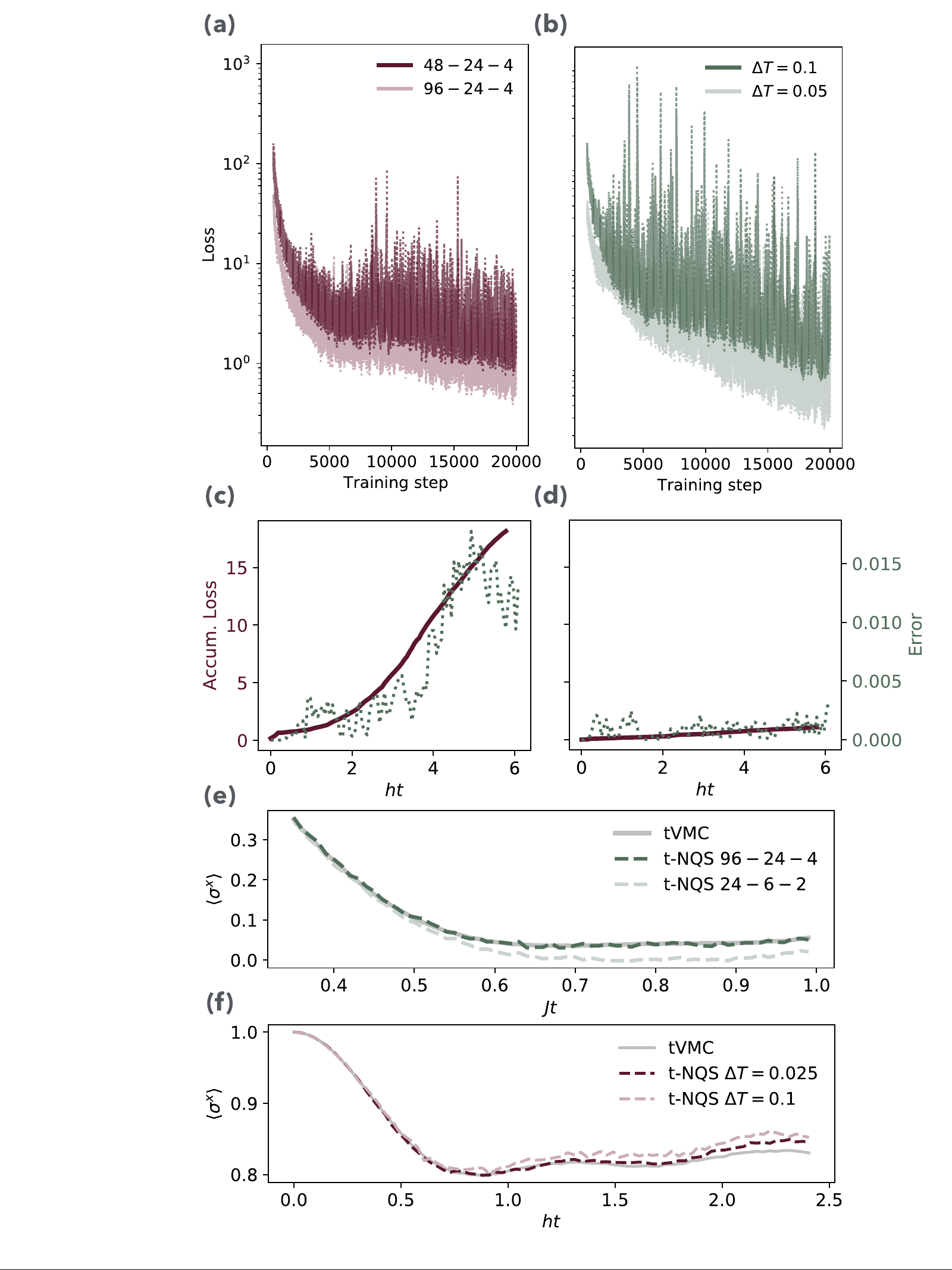}
\centering
\caption{(a-b) Performance of the t-NQS model during optimization: (a) Training loss for different Transformer model sizes, demonstrating that larger models achieve lower loss values. Model size is described by $d_e$ - $n_h$ - $n_l$. (b) Training loss for varying time interval lengths $\Delta T = t_1 - t_0$ with the same t-NQS model, showing that shorter intervals, corresponding to more parameters over the same total time, result in lower loss and higher accuracy. (c-d) Relationship between the accumulated loss, $\frac{\Delta t}{T}\sum_{t_i \leq t} \varepsilon(t_i)$, and the error in $\langle \sigma^x \rangle$, defined as $|\langle \sigma^x \rangle_{\text{MPS}} - \langle \sigma^x \rangle_{\text{t-NQS}}|$, for TFI quenches with PBC at (c) $h = h_c$ and (d) $h = 2h_c$. (e-f) Comparison of TFI with PBC quench results: (e) for $h = 0.1h_c$ with small and large model sizes, and (f) for $h = h_c$ with long and short time intervals, illustrating how model size and time interval act as fine-tuning mechanisms to enhance accuracy by increasing the number of trainable parameters. }
\label{fig:loss}
\end{figure}

Fig.~\ref{fig:loss}a displays the loss \eqref{eq:cost_function} during optimization for the quench to $h=h_c$ with two different sizes of the transformer t-NQS. The t-NQS size is changed by varying the embedding size of different layers as indicated in the figure legend; these sizes correspond to $10^5$ and $4 \cdot 10^5$ parameters, respectively. Increasing the t-NQS size clearly reduces the reached cost. 

When working with concatenated time intervals, the interval length $\Delta T$ can in fact serve as an alternative parameter, that controls the effective t-NQS size. Solutions on longer intervals are more complex and therefore require larger t-NQS. Accordingly, we find in Fig.~\ref{fig:loss}b, that reducing the interval length leads to a reduction of the final cost. Importantly, the accumulated training loss is closely tied to the accuracy of the results for observables, as shown in Fig.\ref{fig:loss}c and d. We observe, that deviations in the observables exhibit a nearly linear correlation with the accumulated residuals defined in Eq.~\eqref{eq:accumulated_residuals}. 
This highlights that monitoring the accumulated residuals serves as a reliable indicator of the model's accuracy.  

In addition to providing flexibility in tuning the model's performance, these results establish a direct connection between model size, time interval, and accuracy. As shown in Fig.\ref{fig:loss}e and f, increasing the number of trainable parameters and reducing the time interval significantly enhances the accuracy of the observables. Systematically adjusting the time interval while scaling the model size enables precise control over the trade-off between computational cost and accuracy.



\subsection{Extrapolation to unseen times}
\begin{figure}[t]
\includegraphics[clip, trim=11cm 4cm 16.5cm 6cm, width=8.5cm]{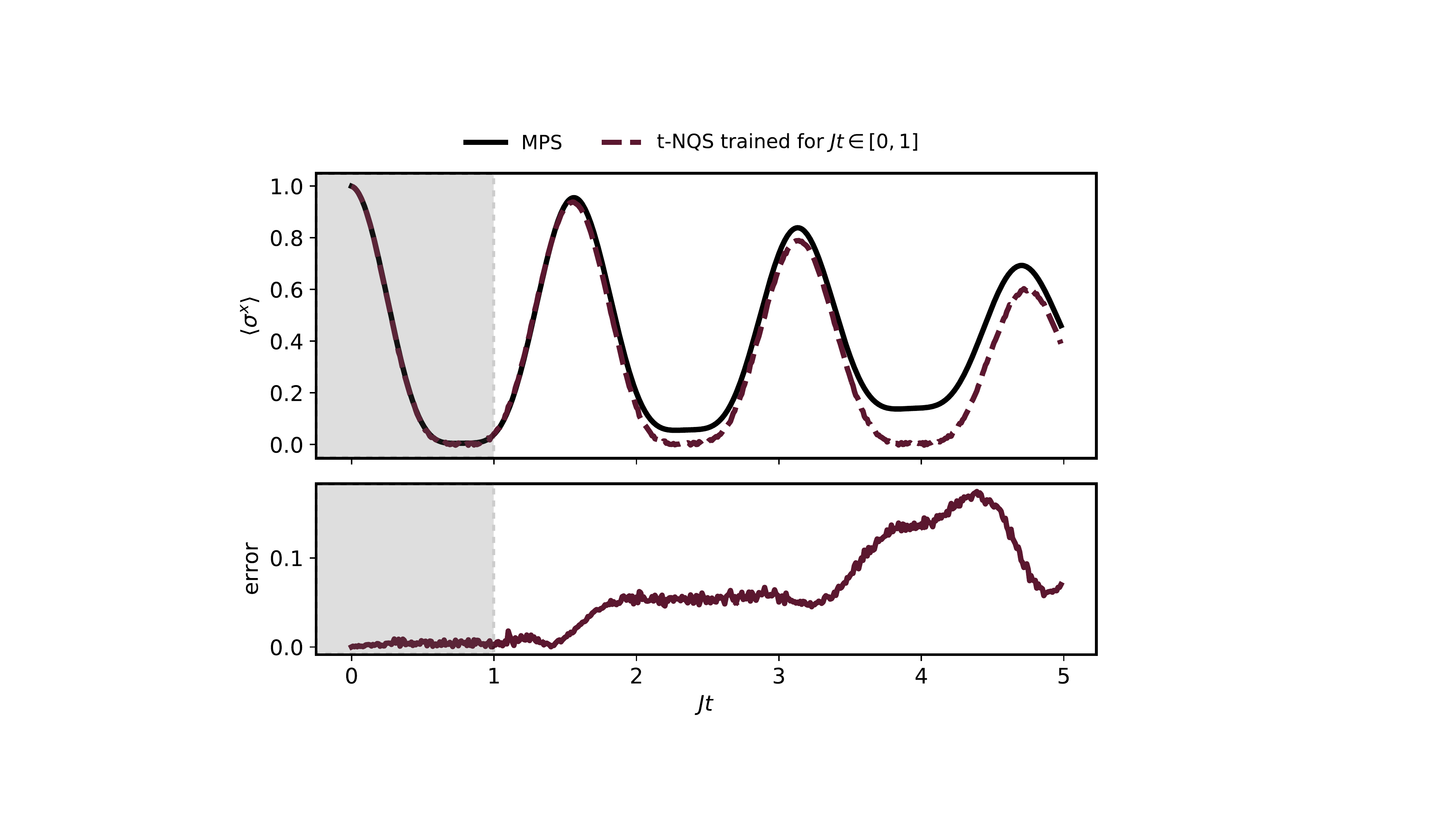}
\centering
\caption{Magnetization dynamics of a 4×4 system following a TFI quench with PBC, starting from the initial state $\ket{\psi_0} = \ket{\rightarrow,...,\rightarrow}$. The error in the lower panel is defined as $|\langle \sigma^x \rangle_{\text{MPS}} - \langle \sigma^x \rangle_{\text{t-NQS}} |$. The t-NQS transformer model was trained on the interval $Jt \in [0, 1]$, yet demonstrates excellent agreement with time-dependent MPS results over a significantly extended time scale. The results are presented up to $Jt=5$, demonstrating the model's ability to generalize well beyond its training range. }
\label{fig:pred}
\end{figure}

Additionally, we show that the t-NQS model can transfer knowledge to new time intervals it has never seen before. To demonstrate this, we consider a 2D TFIM with $4 \times 4$ sites and first train a t-NQS model for a time interval $Jt \in [0,1]$. As a reference, we use time-dependent MPS results obtained using TDVP. Considering the transverse magnetization $\langle \sigma^x \rangle$, the t-NQS shows very good agreement with time-dependent MPS results within this interval, see Fig.~\ref{fig:pred}. Without any additional training, we let the model \textit{predict} the quantum state at later time steps. In the vicinity of the training region, the model continues to show very good agreement for later time steps, even though it has not trained in this region. Extending the prediction to significantly longer time scales compared to the training interval, the t-NQS still captures the significant features, although the quantitative agreement deteriorates. This shows the ability of the t-NQS model to extrapolate to regions it has not trained upon and thus without any additional computational cost, however with slightly larger error.

\section{Conclusion and outlook}

In this work, we introduced the time-dependent neural quantum state (t-NQS) as a novel framework for simulating quantum dynamics. The central innovation of t-NQS lies in its explicit time-dependent ansatz, which enables the optimization of a single, time-independent set of parameters $\theta$ to solve the time-dependent Schrödinger equation across an entire time interval. This approach reformulates quantum dynamics as a global optimization problem, eliminating the need for traditional step-by-step integration methods. Instead, the model learns to represent the dynamics directly through an efficient autoregressive transformer architecture, where time is treated as an input parameter. 
Our results on the transverse field Ising model and the Heisenberg model with a time-dependent staggered field illustrate the accuracy, scalability, and strength of this method in capturing the intricate dynamics of quantum systems. 

One of the most compelling features of this approach is its predictive capability. Once trained, the t-NQS can extrapolate the dynamics to time intervals beyond those seen during training, showcasing its ability to generalize and infer temporal behavior in a way that traditional methods cannot. 
We conjecture, that this generalization capability is also the reason for a substantially enhanced sample efficients of the t-NQS compared to other time-dependent NQS algorithms, because information can be shared between different time points in the global variational approach. Conventional approaches require of the order of $10^5$ samples to propagate the wave function for one time step, whereas we find less than 100 samples sufficient for comparable step sizes, see Methods.

The results presented in this work were obtained with the rather modest computational resources of a single NVIDIA H100 GPU. The optimization algorithm is, however, highly parallelizable, because the computationally most demanding part of ANN evaluations can be straightforwardly distributed across multiple devices. We therefore expect, that the approach can be scaled up to substantially larger system and NQS sizes. The formulation of time-dependent simulations as a global optimization problem may moreover open new avenues towards foundation models for quantum dynamics analogous to the first steps taken in that direction for ground state problems \cite{fitzek2024rydberggpt, zhang2023transformer}.

\begin{acknowledgments}
We gratefully acknowledge discussions with Tatiana Vovk, Teoh Yi Hong, Hannah Lange, Roeland Wiersema, Schuyler Moss, and Roger Melko. We thank Pit Neitemeier and Dante Kennes for sharing their unpublished work. We acknowledge support by the Deutsche Forschungsgemeinschaft (DFG, German Research Foundation) under Germany’s Excellence Strategy—EXC-2111—390814868.
MS was supported through the Helmholtz Initiative and Networking Fund, Grant No.~VH-NG-1711. \\

\textit{Note added:} During the preparation of this manuscript, we became aware of a related work that has been carried on in parallel by A. Sinibaldi, D. Hendry, F. Vicentini and G. Carleo, which will appear simultaneously on the preprint server.
\end{acknowledgments}


\section*{Methods}

\paragraph*{Computational complexity.}

Here, we outline the computational complexity of the algorithm proposed in this paper. The computation of the cost function consists of different parts:  

A \textit{forward pass} to obtain $\psi_\theta(\sigma, t)$: The complexity of the computation of the output of a simple encoder and decoder network is dominated by the activation function and the feed forward network, and is primarily determined by the number of tokens in the input sequence $N$, the number of samples $N_s$, and the embedding dimension $d_e$. The overall complexity of the forward pass with an encoder-decoder transformer is $O(n_l(d_e^2 N N_s N_{ts} + d_e N^2 N_s^2 N_{ts}^2))$.   

 A \textit{backward pass} to obtain $\frac{\partial \psi_\theta}{\partial t}(\sigma, t)$: For each layer, the computational complexity of backpropagation is approximately the same as the forward pass. The storing of the gradients of all the parameters requires overhead, which increases memory usage but does not change the computational complexity.   
 
The \textit{sampling algorithm and local energy}: The autoregressive neural network allows for direct, and efficient sampling with a computational complexity of a single forward pass \cite{sharir2020deep}. The computational complexity of evaluating the local energy for a given configuration and timestep $\sigma, t$ is $O(k n_l(d_e^2 N N_s + d_e N^2 N_s^2))$, with $k$ the locality of the Hamiltonian. For local Hamiltonians, $k$ is small (e.g. on the order of the number of nearest neighbors in a lattice), however for global interactions this is larger and the number of required forward passes increases.  
 
The overall computational cost of the optimization is further influenced by the number of training steps and the range of time steps used during training.  

The cost of forward-propagation of an NQS in tVMC is analogously dominated by the necessary NQS evaluations. Typical sample numbers for accurate estimation of the TDVP equation are of the order of $N_\mathrm{MC}=10^5$. Just as t-NQS, tVMC requires one evaluation of the local energy per sample configuration, which introduces a factor of system size $N$. Moreover, the non-linear ODE is typically integrated with adaptive time steps, which comes at the cost of sampling at least five times per simulation time step. Typical time steps are of the order of $10^{-3}$ in the natural units, i.e., comparable to the values we found suited for t-NQS, see Table \ref{tab:param}. Overall, this means, that t-NQS -- in our case using at most $N_s=100$ samples per time point -- is two to three orders of magnitude more sample efficient than the state of the art tVMC approaches for NQS.




\begin{table}[htbp]
\centering
\label{tab:param}
\renewcommand{\arraystretch}{1.5} 
\resizebox{\columnwidth}{!}{%
\begin{tabular}{lccccccc} 

  & $N_s$ & $\delta t$ & $\Delta T$ & $N_{ts}$ & $n_l$ & $n_h$ & $d_e$ \\  \hline 
Fig.\ref{fig:mainFig}(c)          & 20             & 0.01                & 1                   & 100               & 2             & 6              & 24             \\ 
Fig.\ref{fig:results}(a)           & 2              & 0.001               & 0.05                & 50                & 4             & 24             & 96                   \\ 
Fig.\ref{fig:results}(b)          & 2              & 0.0005              & 0.025               & 25                & 4             & 24             & 96      \\ 
Fig.\ref{fig:results}(c)          & 10             & 0.01                & 0.2                 & 20                & 4             & 24             & 96             \\ 
Fig.\ref{fig:pred}                & 100            & 0.01                & 1                   & 100               & 4             & 24             & 96             \\ 
\end{tabular}%
}

\caption{
Transformer and training parameters used for each result. Here, $N_s$ is the number of samples we train per time step, and $N_{ts}$ is the number of time steps trained in each training step. The total time interval of the t-NQS, $\Delta T = t_1 - t_0$, is divided into smaller time steps of size $\delta t$. Other parameters include $n_l$, the number of layers; $n_h$, the number of attention heads; and $d_e$, the embedding dimension. }
\label{tab:param}
\end{table}

\paragraph*{Optimization.}

For optimization, we utilize the Adam algorithm, which enhances stochastic gradient descent by introducing adaptive learning rates and regularization techniques. 
Additionally, we integrate a learning rate scheduler to further enhance training stability and convergence. 
A fixed learning rate may lead to suboptimal training, as a high learning rate can cause instability, while a low learning rate can result in slow convergence. The scheduler dynamically adjusts the learning rate to balance these issues. In this work, we vary the learning rate according to the formula:
\begin{equation}
     \text{lr}(i_{\text{step}}) = 5 d_e^{-0.5} \text{min}(i_{\text{step}}^{-0.75}, i_{\text{step}} i_{\text{warm-up}}^{-1.75}),
\end{equation}
which gradually decreases the learning rate. This scheduler is taken from Ref.~\cite{zhang2023transformer}. We use $i_{\text{warm-up}} = 4000$, which corresponds in increasing the learning rate during the first $4000$ steps, after which the learning rate is polynomially decreased. 

The further training hyperparameters are summarized in Table \ref{tab:param}.

\paragraph*{Transformer architecture}
The most important element in the transformer architecture is the self-attention mechanism, which encodes correlations between different input elements. Masked self-attention has proven effective in capturing long-range correlations - including entanglement - present in quantum systems \cite{sprague2024variational}. It provides trained connections to all previous elements in the sequence, without requiring the sequential transfer of hidden or latent vectors \cite{vaswani2017attention,lange2024architectures}. The scaled dot-product attention connects input query ($Q$) and key ($K$) matrices with an input value matrix ($V$) as
\begin{equation} \label{eq:Attention}
    \text{Attention}(Q,K,V) = \text{Softmax}\left(\frac{QK^T}{\sqrt{d_e}} + M\right) V
\end{equation}
where $\frac{1}{\sqrt{d_e}}$ is a scaling factor dependent on the embedding dimension. These query, key and value matrices are obtained with different trainable, linear mappings of the embedded input vector. Here we further use multi-headed attention, as discussed in \cite{vaswani2017attention, sprague2024variational}, where the number of heads is given by $n_h$. This self-attention layer generates all-to-all interactions between all the sites in the system. In the decoder, a masking term $M$ ensures that only connections between sites $\sigma_i$ and $\sigma_{<i}$ are allowed by setting all other connections to $-\infty$, to preserve the autoregressive property. Subsequently, an activation function maps the output of the masked attention mechanism to an encoded input vector. We adopt the transformer model as proposed in Ref. \cite{vaswani2017attention}, where a second, mixed masked attention mechanism combines the context vector, which captures the time, with this encoded input vector, that captures the spin configuration, see Fig.\ref{fig:mainFig}a. In this instance, the attention mechanism maps the context vector to the query and key matrices, and the encoded input vector into the value matrix, which in turn are connected with Eq. \eqref{eq:Attention} and passed through a FFNN and different linear layers to the output amplitude and phase. We employ regularization techniques such as dropout and layer normalization as has been proposed in Ref. \cite{vaswani2017attention}. The complete decoder cell, encompassing the masked attention mechanism, the mixed attention mechanism, the feed-forward neural network layer, and the add-and-norm operations, can be repeated multiple times to enhance the model's expressiveness. This repetition is defined by the number of layers $n_l$. 

For the decoder we adopt an autoregressive model, which learns the conditional probability at a certain time step $t$ of said spin state  $p(\sigma_i | \sigma_{i-1}, ..., \sigma_{1}, t)$. The joint probability distribution at this time step is then given by 
\begin{equation}
    p_\theta(\sigma, t) = \prod_{i=1}^N p_\theta(\sigma_i | \sigma_{i-1}, ..., \sigma_1, t).
\end{equation}
The autoregressive structure enables efficient sampling, unlike energy-based neural network models. By normalizing each conditional probability $ p_\theta(\sigma_i | \sigma_{<i}, t) $, the resulting output probability is inherently normalized. This is done with a Softmax activation function. The Softmax function is applied after the initialization layer, to initialize the model at $t=t_0$ while simultaneously normalizing the NQS model for each time step. This approach allows for direct sampling from the amplitudes, obtaining independent and identically distributed samples, avoiding the more complex procedures, such as Markov chain sampling, required for non-autoregressive architectures. 

We model the complex-valued quantum state at a certain point in time with the neural network. We can relate the output probability distribution of the transformer network to the amplitude of the quantum state through Born's rule:
\begin{equation}
    p_\theta(\sigma, t) = |\psi_\theta(\sigma, t)|^2 = a_\theta(\sigma, t)^2.
\end{equation}
We construct the phase in a similar autoregressive fashion:
\begin{equation}
    \phi_\theta(\sigma, t) = \prod_{i=1}^N \phi_\theta(\sigma_i | \sigma_{i-1}, ..., \sigma_1, t).
\end{equation}

The autoregressive sampling avoids the need for Markov Chain Monte Carlo (MCMC), eliminating potential autocorrelation issues and ensuring efficient, unbiased sampling.

The initialization layer of the t-NQS network is defined by a linear combination of the amplitude and phase of the state, as shown in Eq.\ref{eq:concat}, with a function $f(t)$ that adheres to the conditions $f(t_0) = 1$ and $f(t_1)=0$. To achieve this, we opt for an exponentially decreasing function, while allowing for a rapid increase in the neural network's contribution:
\begin{equation}
    f(t) = e^{-n\frac{t-t_0}{t_1-t_0}} - e^{-n}\frac{t-t_0}{t_1-t_0}
\end{equation}
with $n \in \mathbb{R}^+$ an initialization parameter. 

The computational complexity of the t-NQS method primarily arises from the optimization of the neural network parameters. For an autoregressive, attention-based transformer architecture, the computational cost per optimization step is dominated by the self-attention mechanism, which typically scales as $O(n_l(N^2 N_s^2 N_{ts}^2 d_e + N N_s N_{ts} d_e^2 ))$. With $N_{ts}$ the number of included target time steps, $N_s$ the number of samples, and $n_l$ the number of layers in the decoder. This method generally requires very few samples; the results presented are achieved with fewer than 20 samples per time step per training step, resulting in the method's high efficiency compared to other established NQS approaches. All the results presented were obtained on a single GPU; however, the method allows for straightforward parallelization across multiple devices.

\bibliography{main}

\end{document}